\documentclass[twocolumn,showpacs,preprintnumbers,amsmath,amssymb]{revtex4}

\usepackage{graphicx}
\usepackage{rotating}
\usepackage{dcolumn}
\usepackage{bm}

\begin{document}
\date{\today}
\title{Entropy driven atomic motion in laser-excited bismuth}
\author{Y. Giret$^{1}$, A. Gell\'e$^{1}$ and B.~Arnaud$^{1}$}
\affiliation{$^{1}$Institut de Physique de Rennes (IPR),
UMR UR1-CNRS 6251, Campus de Beaulieu - Bat 11 A, 35042 Rennes Cedex, France, EU}
\date{\today}

\begin{abstract}

We introduce a thermodynamical model based on the two-temperature approach in order to 
fully understand the dynamics of the coherent A$_{1g}$ phonon in laser-excited bismuth.
Using this model, we simulate the time evolution of (111) Bragg peak intensities measured by Fritz {\it{ et al}}~ 
[Science {\bf 315}, 633 (2007)] in femtosecond X-ray diffraction experiments performed on a bismuth film
for different laser fluences. The agreement between theoretical and experimental results is striking
not only because we use fluences very close to the experimental ones but also because 
most of the model parameters are obtained from {\it{ab-initio}} calculations performed 
for different electron temperatures. 

\end{abstract}
\maketitle

The development of optical pump-probe experiments has revolutionised physics
by allowing the study of non-equilibrium processes on a femtosecond
time scale. Indeed, an ultrashort laser pulse can cause dramatic changes in the
physical properties of covalently bonded solids and induce ultrafast melting,
phase transitions or coherent phonons\cite{bennemann_2004}. Coherent excitation
of optical phonons is a general phenomenon which has been extensively investigated
in bismuth where the reflectivity change following electronic excitations exhibits
damped oscillations arising from atomic motions corresponding 
to the A$_{1g}$ mode\cite{Zeiger_1992}. 
Unfortunately, it is not possible to directly relate the amplitude of the reflectivity
signal to the atomic positions. The goal of measuring atomic position as a function
of time remained elusive until the breakthrough of time resolved X-ray diffraction
experiments which allow atomic motions to be followed stroboscopically with picometer
spatial and femtosecond temporal resolution\cite{pfeifer_2006}. Different femtosecond
X-ray diffraction experiments have been conducted to probe the dynamics of the
A$_{1g}$ phonon in laser-excited Bi\cite{sokolowski_2003, fritz_2007, johnson_2008, johnson_2009}.
In this letter, we try to understand and to reproduce the time evolution of (111) Bragg
peak intensities measured by Fritz {\it{et al}}\cite{fritz_2007}.

Bismuth is a semi-metallic solid, whose structure can be derived from the face-centered-cubic lattice
by a weak rhombohedral distorsion of the cubic unit cell. The crystal basis consists
of two atoms located along the body diagonal of length $c$ at $\pm uc$, where 
$u$ is the A$_{1g}$ phonon coordinate. At 300 K, the equilibrium value of $u$ is
$u_{eq}\simeq$ 0.234\cite{schiferl_1969}. It is now well
established that the sudden change of $u_{eq}$ following laser excitation explains the
generation of coherent phonons in Bi. The time evolution of $u$ can be obtained
from the displacive excitation of coherent phonons (DECP) model\cite{Zeiger_1992}.
This model, based on the idea that $u_{eq}$ increases linearly as
a function of the excited carriers density $n$ and that $n$ decreases exponentially
as a function of time, however does not capture all aspects of the phonon and electron dynamics.
Two approaches based on {\it{ab-initio}} calculations\cite{murray_2005, zijlstra_2006} have
been proposed to describe the non-equilibrium electron distribution.
Murray {\it{et al}}\cite{murray_2005} assumed that electrons and holes can be described by
two Fermi-Dirac (FD) distributions with the same temperature but different chemical potentials, and obtained
a good agreement with experiments concerning the early oscillations 
of the A$_{1g}$ phonon\cite{fritz_2007}.
Zijlstra {\it{et al}}\cite{zijlstra_2006} studied the coupling between the A$_{1g}$ and E$_g$ modes and 
assumed that electrons and holes
can be described by a unique FD distribution. They claimed that a two-chemical potential
model does not provide a realistic description of the electron dynamics in Bi, at least for high fluences.
While such a claim seems to be supported by recent work based on a two-fluid model\cite{johnson_2008}, the
relevance of the two approaches remains unclear\cite{zijlstra_2010, johnson_2010}.

In this work, we consider a model based on the two-temperature approach\cite{kaganov_1957} 
where the electron system is locally described by a single FD distribution, 
the A$_{1g}$ phonon mode obeys a classical equation of motion and all the remaining modes
are lumped in a unique phonon bath with specific heat $C_l$. We assume that both the electron temperature
$T_e$ and lattice temperature $T_l$ are well defined at each time because of electron-electron and phonon-phonon
interactions.
Making a local energy balance, the heat respectively received by the electron and phonon systems is given by:
\begin{equation}
\label{eq_heat_e_S}
T_e\frac{\partial S}{\partial t}= \frac{\partial U}{\partial t}
+\frac{f c}{v}\frac{\partial u}{\partial t}= P+\frac{\partial}{\partial
  z}\left(\kappa_e \frac{\partial T_e}{\partial
  z}\right)-G_0\left(T_e-T_l\right)
\end{equation}
\begin{equation}
\label{eq_heat_r}
C_l\, \frac{\partial T_l}{\partial t}= +G_0\left(T_e-T_l\right)
+\frac{4Mc^2}{v\tau}\left( \frac{\partial u}{\partial t}\right)^2
\end{equation}
where $P$ is the energy deposited by the laser pulse in the electron system per unit
volume and time, $S$ and $U$ are electronic entropy and energy per unit
volume, $f$ is the force acting on the A$_{1g}$ phonon, $\kappa_e$ is the
electron thermal conductivity, $G_0$ is the electron-phonon coupling constant which
governs the heat transfer from the electron system to the lattice,
$v$ is the unit cell volume, 
$M$ is the mass of a Bi atom and $\tau$ is the lifetime of the A$_{1g}$ phonon. 
Since the laser spot diameter is much larger than the penetration depth we only
consider the spatial dependence of the parameters in the direction
perpendicular to the surface, labelled as $z$. One should notice that all
parameters of Eq.~\ref{eq_heat_e_S} and~\ref{eq_heat_r} depend on $T_e$ and
$u$, both of which depend on the depth $z$. These equations have to be solved
together with the phonon equation of motion~:
\begin{equation}
\label{eq_motion}
\frac{\partial^2 u}{\partial t^2}= \frac{f}{2Mc}-\frac{2}{\tau}
\frac{\partial u}{\partial t}
\end{equation}
where the last term describes the energy transfer from the coherent phonon 
to the incoherent phonon bath, and where the force $f$ is defined by: 
\begin{equation}
f= -\frac{v}{c}\left. \frac{\partial U}{\partial u}\right|_{S}.
\end{equation}


In order to evaluate the force $f$ acting on the coherent phonon and other quantities involved in our model, 
we performed {\it ab-initio} calculations for different electron temperatures $T_e$ and phonon coordinates $u$.
All these calculations were done within the framework of the local density approximation (LDA) for
the exchange-correlation functional to density functional theory (DFT) 
using the ABINIT code\cite{gonze_2009}. Spin-orbit coupling was included and an energy cut-off of
15 Hartree in the planewave expansion of wavefunctions together with a $12\times 12\times 12$ kpoint grid for the
Brillouin zone integration were used. These parameters ensure the convergence of both vibrational and 
electronic properties\cite{diaz_2007}. The experimental lattice parameters
at room temperature 
are used\cite{schiferl_1969} and volume changes due to an increase of the lattice temperature $T_l$ are neglected. 

\begin{figure}
\vskip1.0truecm
\includegraphics[width=8.5cm]{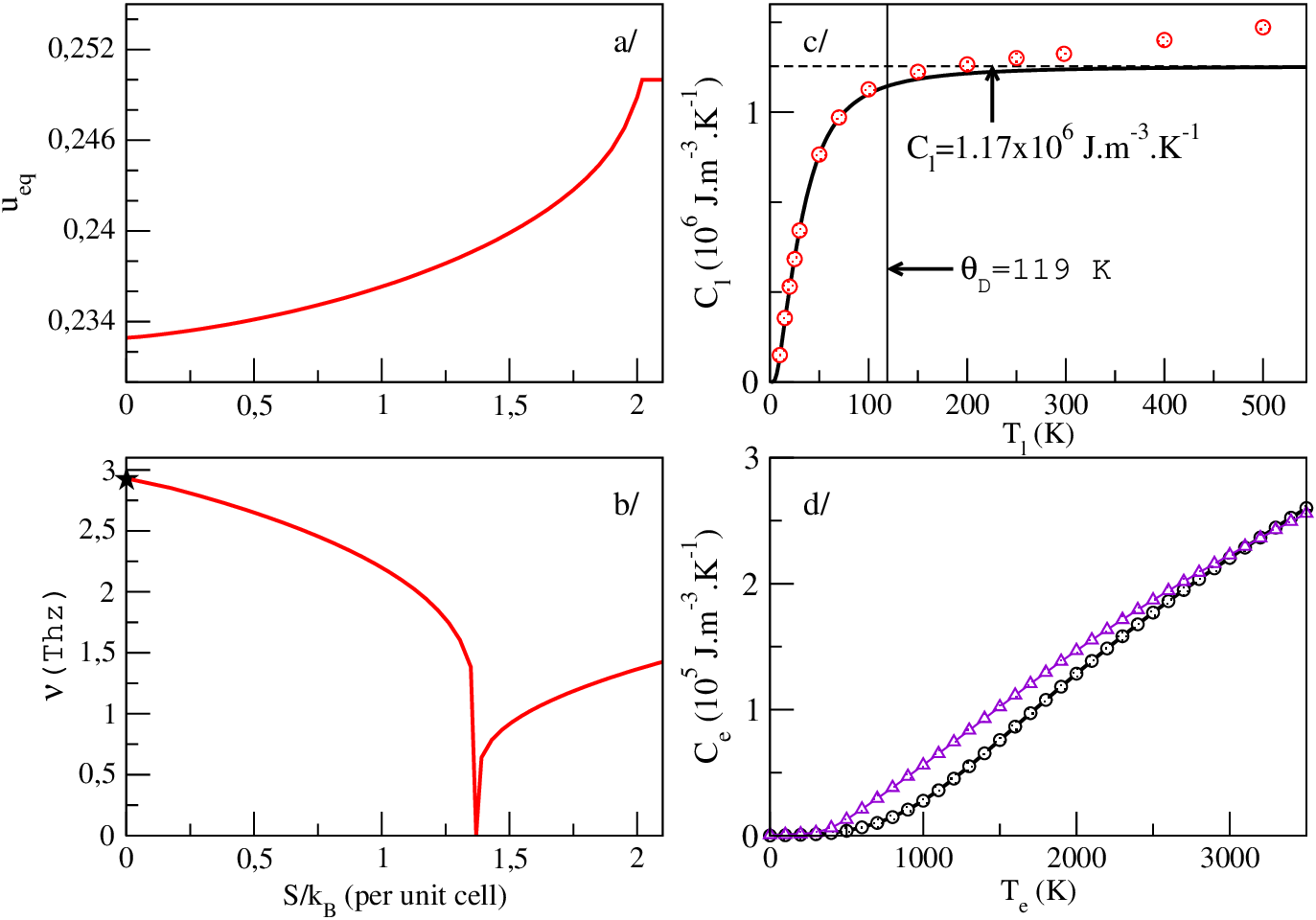}
\caption{\label{ab-initio-quantities}
(color online) a/ $u_{eq}$ as a function of electronic entropy $S$.
b/ A$_{1g}$ phonon frequency $\nu$ (in Thz) as a function of $S$. The star indicates the Raman frequency measured
at 300 K\cite{handbook}.
c/ The calculated constant-volume lattice specific heat $C_l$ (solid line) of Bi compared to experimental data
(open circles) from Ref. \cite{handbook}
for lattice temperatures $T_l$ up to the melting temperature $T_f=544$ K.
d/ Calculated electron specific heat $C_e$ as a function of electron temperature $T_e$ for 
$u$=0.233 (black circles) and $u$=0.24 (violet triangles).
}
\end{figure}

First, we calculated both energy $U$ and entropy $S$ as a function of $u$ 
and $T_e$ and hence obtained $U(S,u)$. 
Figure \ref{ab-initio-quantities}.a shows that the 
equilibrium position $u_{eq}$ increases from 0.233 to 0.25 as $S$ increases from 0 to $\sim 2.0$ k$_B$ per
unit cell. This finding confirms the displacive mechanism of phonon generation and is qualitatively
in line with the behaviour observed by Murray {\it et al}\cite{murray_2005} using a two chemical potential 
approach or by Zijlstra {\it et al}\cite{zijlstra_2006} using the same type of approach as in the work reported here.
If we assume that the laser pulse energy is deposited homogeneously at the end of the pulse ($\kappa_e\to\infty$),
Eq. \ref{eq_heat_e_S} shows that the electronic entropy $S$ becomes a constant of motion provided that the energy
transfer from the electron system to the lattice is frozen ($G_0=0$). The solution of Eq. \ref{eq_motion}
when $\tau\to\infty$ (undamped dynamics) then provides the phonon frequency $\nu$ of the A$_{1g}$ mode 
for a given entropy $S$ which depends on the laser pulse fluence.
Figure \ref{ab-initio-quantities}.b shows that $\nu$
decreases from 2.93 Thz to 0 Thz when $S$ increases
from 0 to $\sim 1.37$ k$_B$ per unit cell. This red shift is still in qualitative agreement 
with previous theoretical studies\cite{murray_2005, zijlstra_2006}. 

We have also calculated the lattice specific heat $C_l$ within the harmonic 
approximation\cite{cardona_2007}.
As shown in Fig. \ref{ab-initio-quantities}.c, 
the agreement between our calculated lattice specific heat and the experimental data\cite{handbook} 
is good up to $T_l\sim 300$ K. The discrepancies between theoretical and experimental results become
larger at higher temperature because of anharmonic interactions neglected in our calculation. In our
simulations, the initial lattice temperature $T^0$ was set to 300 K.
As $T^0\gg\theta_D$ where $\theta_D=119$ K is the Debye temperature, the temperature
dependence of $C_l$ can be neglected. Indeed, $C_l(T^0)$ is only 1 \% smaller than the value
given by the Dulong-Petit law shown as a dashed line in Fig. \ref{ab-initio-quantities}.c.
The electron temperature dependence of $C_l$ can also be safely neglected.


Finally, we calculated the electron specific heat of Bi, $C_e=\partial U/\partial T_e$, as a function 
of electron temperature $T_e$ for different values of $u$. As expected, the
temperature dependence of $C_e$ is linear at low electron temperatures.
Figure \ref{ab-initio-quantities}.d shows that a significant positive
deviation from the linear behaviour occurs when $T_e$ becomes larger than $\sim 300$ K and that this
deviation increases when $u$ increases, i.e. when Bi goes toward the hypothetical metallic state corresponding
to $u=0.25$. The knowledge of $C_e(T_e)$ is crucial, not only because it gives a correct estimate of
the rise of temperature in the electron system, but also because it is related to the electron thermal conductivity
$\kappa_e$. Indeed, assuming that the total electron scattering rate is dominated by the electron-phonon
scattering rate, the electron thermal conductivity can be approximated by 
$\kappa_e(T_e, T_l)=\kappa_0\times (C_e(T_e)/C_e(T^{0}))\times(T^{0}/T_l)$\cite{lin_2008, kanavin_1998}
where the dependence on $u$ has been omitted and where  
$\kappa_0=$ 11 W.m$^{-1}$.K$^{-1}$ is the experimental value at room temperature\cite{gallo_1963}.


\begin{figure}[htb]
\resizebox{7.5cm}{!}{\includegraphics*{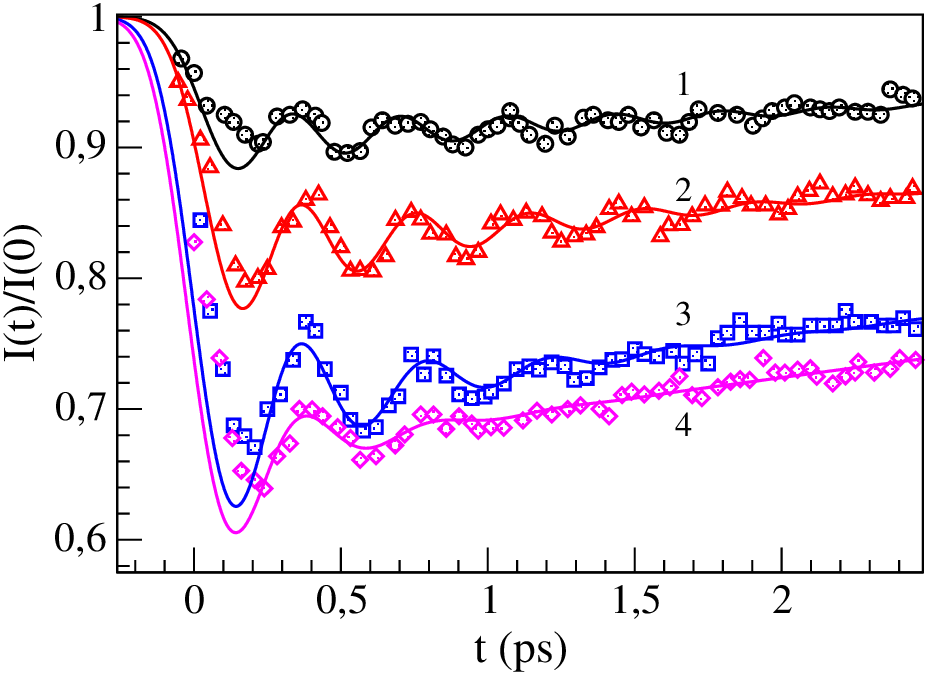}}
\caption{(color online) Normalized intensities of the (111) Bragg peak
measured by Fritz {\it{ et al}}\cite{fritz_2007} as a function of time delay $t$ between the laser pulse and
X-ray probe for different absorbed fluences (symbols) compared to the theoretical calculation (solid lines).
The measured fluences are 0.7 (black circles), 1.2 (red triangles), 1.7 (blue squares) 
and 2.3 mJ.cm$^{-2}$ (magenta losanges).
The parameters used in our simulations are given from top to  bottom for curves labelled 1 to 4: 
the absorbed fluences $F_{abs}$ are 
0.64 (black), 1.26 (red), 2.18 (blue) and 2.47 mJ.cm$^{-2}$ (magenta);
The effective electron-phonon coupling constants $G_0$ are 0.84, 1.15, 1.40 and 1.45$\times 10^{16}$ 
W.m$^{-3}$.K$^{-1}$ and the decay times $\tau$ are 0.87, 0.64, 0.44 and 0.25 ps respectively.
The uncertaintities for $\tau$ are quite large for curves 1 and 2 because of the limited duration of the
experiments. The fitted value of the FWHM of the X-ray pulses is $t_X=205$  fs.}
\label{f_dyn}
\end{figure}
We can now simulate 
the time-resolved X-ray diffraction experiments performed by Fritz {\it et al}\cite{fritz_2007} on
a Bi film of thickness $L=$ 50 nm excited by a near-infrared laser pulse whose full width at half maximum
 (FWHM) is $t_w=70$ fs. The source term $P(z,t)$ in Eq.~\ref{eq_heat_e_S}
is given by:
\begin{equation}\label{source-term}
P(z,t)=\frac{2 F_{abs}}{l_p t_w}\sqrt{\frac{\ln 2}{\pi}} \exp\left[-4 \ln 2\,\frac{t^2}{t_w^2}\right]
\exp\left[-\frac{z}{l_p}\right]
\end{equation}
where $l_p=14$ nm is the penetration depth at wavelength of 800 nm\cite{handbook} and $F_{abs}$ 
is the absorbed fluence. 
The solution of the coupled differential equations \ref{eq_heat_e_S}, \ref{eq_heat_r} and \ref{eq_motion} 
for a set of the two unknown physical parameters $G_0$ and $\tau$, which are assumed to remain constant
on the time scale of the experiment for a given fluence, provides
$u(z, t)$ and $T_e(z,t)$. As will be seen later, the spatial dependence of $u$ and
$T_e$ can be neglected only $\sim$ 100 fs after the arrival of the laser pulse on the Bi surface. Therefore,
the normalized intensity of the (111) Bragg peak is given by
$I(t)/I(0)=\cos^2[6\pi u(t)]/\cos^2[6\pi u(0)]$
and is convoluted with a Gaussian with FWHM $t_X$ to account for the temporal resolution of the experiment\cite{fritz_2007}.
The results of a least-squares fit of our model to the experimental data are shown as solid curves
in Fig. \ref{f_dyn} for four absorbed fluences.
The parameters $F_{abs}$, $G_0$, $\tau$ and $t_X$ are given in the caption.
The absorbed fluences used in our calculations are very close to the experimental fluences with
the largest deviation occuring for the curve labelled 3 in Fig. \ref{f_dyn}. The
electron-phonon coupling constant $G_0$ increases from 0.84 to 1.45$\times 10^{16}$ W.m$^{-3}$.K$^{-1}$ 
and the decay time $\tau$ of the coherent phonon decreases from 0.87 to 0.25 ps as the theoretical 
fluence increases from 0.64 to 2.47 mJ.cm$^{-2}$. The reported electron-phonon coupling constants are small compared
to usual values in metals\cite{lin_2008} because the density of states at the Fermi level is very small for Bi. 
Figure \ref{f_dyn} shows that the oscillations of the (111) Bragg intensities are nicely reproduced 
by our quasi-isentropic model with the exception of curve 3, where $F_{abs}$ is overestimated with respect to
the experimental fluence. It is worth remarking that curves 3 and 4 are very close to each other while the 
experimental fluences differ by 0.6 mJ.cm$^{-2}$. Interestingly, an upward shift of curve 3 by about 0.05 leads
to a better agreement with our theoretical intensity for a fluence $F_{abs}$=1.79 mJ.cm$^{-2}$ very close 
to the experimental one.
\begin{figure}
\vskip1.0truecm
\includegraphics[width=7.5cm]{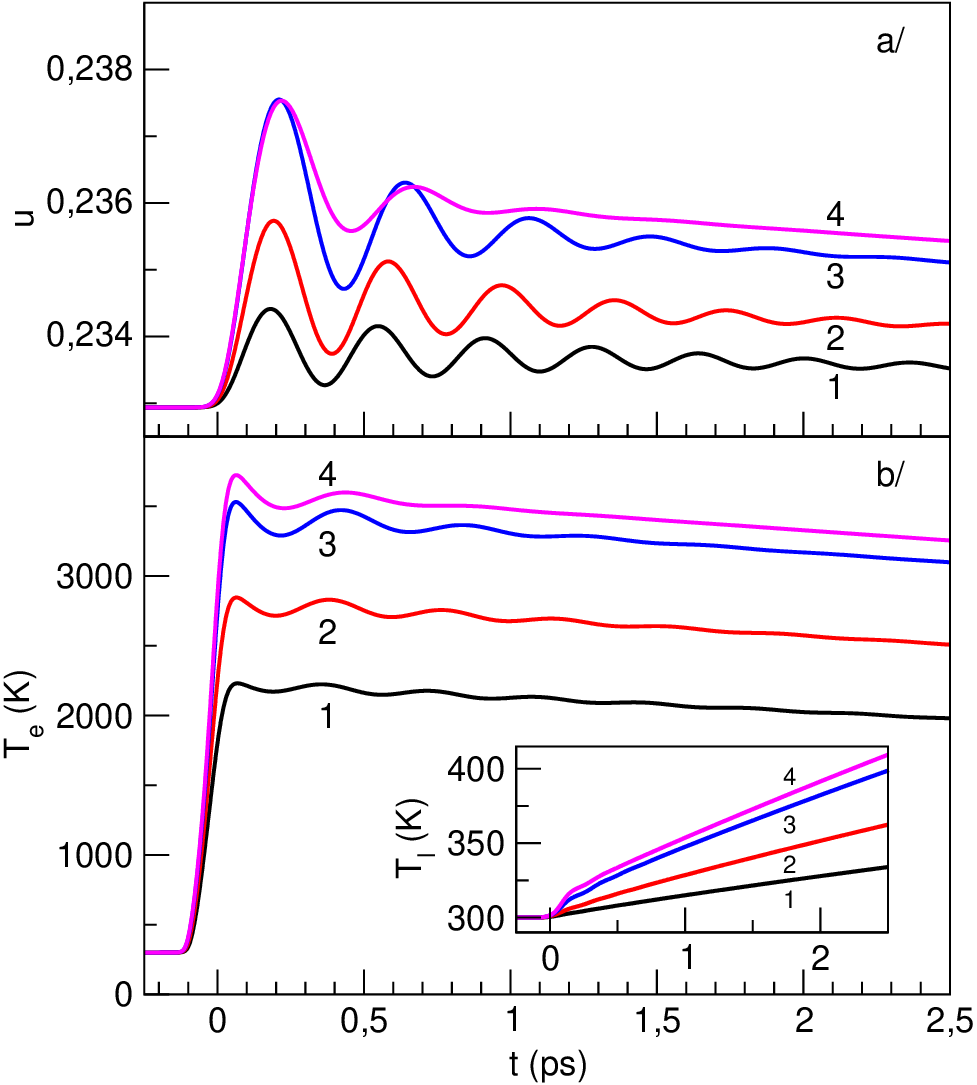}
\caption{\label{u-and-Te-vs-time}
(color online) a/$u$ as a function of time delay $t$ for the parameters given in the
caption of Fig. \ref{f_dyn}.
b/Electron temperature $T_e$ as a function of time delay $t$. 
The inset displays the variation of the lattice temperature $T_l$ as a function of $t$.
The curves labelled 1 to 4 are related to the curves with the same label in Fig. \ref{f_dyn}.
}
\end{figure}

One can now discuss the physical processes playing a role in the generation of coherent phonons. The
laser energy is initially deposited in the electron system since the laser pulse duration
is much shorter than the characteristic time for the electron-lattice energy exchange. At the same
time, the heat diffuses in the electron system. The diffusivity $\chi$ is approximately constant
and given by $\kappa_0/C_e(T^0)\simeq 100$ cm$^2$.s$^{-1}$. Therefore, the usual electron heat
transport occurs and the time needed to uniformize the electron temperature $T_e$ is roughly
given by $L^2/\chi\sim$ 250 fs. Our simulations show that the spatial variations of
$T_e$, $S$ and $u$ can be safely neglected only 100 fs after the laser pulse maximum. 
The strong overheating of the electron system at the end of the laser pulse is illustrated
in Fig. \ref{u-and-Te-vs-time}.b. The electron temperature reaches 2231 K for the lowest fluence
 and 3724 K for the highest fluence. The corresponding increases of electronic
entropy $S$ are respectively 0.39 and 0.98 $k_B$ per unit cell. As shown in Fig. 
\ref{ab-initio-quantities}.a, the increase in entropy leads to an increase in $u_{eq}$. 
The atoms thus start to oscillate around a new equilibrium position which evolves slowly as a function
of time $t$ due to heat transfer from the electron system to the lattice. The damped oscillations
of the reduced coordinate $u$ as a function of time delay $t$ are shown in Fig. \ref{u-and-Te-vs-time}.a.
The initial red-shift of the phonon frequency $\nu$ as the fluence increases is clearly visible
and can be attributed to the decrease of $\nu$ as $S$ increases (see Fig. \ref{ab-initio-quantities}.b). 
Figure \ref{u-and-Te-vs-time} also shows that the oscillations of the phonon coordinate $u$ are 
accompanied by oscillations in the electron temperature $T_e$ whose 
amplitude and damping grow with fluence. The oscillations in $T_e$ and $u$ are in antiphase with respect to
each other. Such a behaviour is easy to understand if one assumes that the electron subsystem undergoes
an isentropic transformation ($G_0\to 0$). As the metallicity of Bi is enhanced when $u$ increases,
the electronic entropy $S$ at constant electron temperature would increase up to a maximum value for the
largest value of $u$. In order to compensate for the increase of $S$, $T_e$ decreases
and reaches its minimum value when $u$ reaches its maximum value. Thus, $T_e$ oscillates at the same 
frequency as $u$ but in antiphase. Interestingly, such a behaviour parallels that of a gas enclosed
in an insulating piston chamber. Assuming that the gas undergoes an isentropic transformation, the piston
when displaced from it's equilibrium position, performs an oscillatory motion while the gas temperature
undergoes a similar oscillatory motion, but in antiphase. In Bi, the electronic
entropy $S$ does not remain constant after the arrival of the laser pulse but decays slowly due to
energy exchange between the electron and lattice subsystems. The inset of Fig. \ref{u-and-Te-vs-time}.b
shows the evolution of the lattice temperature $T_l$ as a function of time delay $t$. On the time scale
of the experiments ($\sim$ 2.5 ps), the increase in $T_l$ ranges from 34 K for 
the lowest fluence to 110 K for the highest fluence. Although the increase of the
lattice temperature is significant for the highest fluence, volume changes are expected to be delayed as 
illustrated by the shift in the Bragg angle occuring on a time scale of $\sim$ 20 ps in experiments\cite{fritz_2007}.
It's thus reasonable to assume that the lattice parameters remain constant in our simulations.

In conclusion, we have developed a thermodynamical model in order to simulate the time
evolution of the A$_{1g}$ phonon coordinate following the arrival of a laser pulse
of a given fluence on a Bi film. The intensities of the (111) Bragg peak measured
by Fritz {\it{ et al}}\cite{fritz_2007} 
are fairly well reproduced by our model for fluences very close to the experimental ones.
This success is noteworthy since the force acting on the coherent phonon as well as most
of the model parameters are obtained from {\it{ab-initio}} calculations.  The only adjustable
parameters are the effective electron-phonon coupling constant $G_0$
and the scattering time $1/\tau$ of the coherent phonon.
Our results show that (1) both parameters increase as the fluence increases, (2) the electronic heat diffusion is
crucial and (3) the oscillations of the coherent phonon are accompanied by an oscillation in the electron
temperature. From a fundamental point of view, this work firmly establishes that a single chemical
potential approach is reliable for describing the excited electrons and provides a complete scenario for
the generation of coherent phonons in Bi films.

This work was performed using HPC resources from GENCI-CINES (Grant 2010-095096). We thank K. Dunseath
and H. Cailleau for useful comments.

\end{document}